\documentclass[aps,prl,twocolumn,superscriptaddress,showpacs]{revtex4}

\usepackage{graphicx}% Include figure files
\usepackage{dcolumn}% Align table columns on decimal point
\usepackage{bm}% bold math

\newcommand{\tc}{$T_{\rm c}$}
\newcommand{\tn}{$T_{\rm N}$}
\newcommand{\td}{$T_{\rm D}$}
\newcommand{\la}{La$_2$CuO$_4$}
\newcommand{\lasr}{La$_{2-x}$Sr$_x$CuO$_4$}
\newcommand{\sr}{Sr$_2$CuO$_2$Cl$_2$}

\newcommand{\cana}{Ca$_{2-x}$Na$_x$CuO$_2$Cl$_2$}
\newcommand{\msr}{$\mu$SR}

\begin{document}

%\preprint{APS/123-QED}

\title{Magnetic Phase Diagram of the Hole-doped \cana\/ Cuprate Superconductor}

\author{K. Ohishi}
 \email{kazuki.ohishi@kek.jp}
\affiliation{
Institute of Materials Structure Science, High Energy Accelerator
Research Organization (KEK), 
Tsukuba, Ibaraki 305-0801, Japan
}
\author{I. Yamada}
\affiliation{
Institute for Chemical Research, Kyoto University, Uji, Kyoto
611-0011, Japan
}
\author{A. Koda}
\author{W. Higemoto}
 \altaffiliation[Present Address: ]{Advanced Science Research
Center, Japan Atomic Energy Research Institute, Tokai, Ibaraki 319-1195,
Japan}
\author{S.R. Saha}
\author{R. Kadono}
 \altaffiliation[Also at ]{School of Mathematical and Physical Science,
The Graduate University for Advanced Studies.}
\affiliation{
Institute of Materials Structure Science, High Energy Accelerator
Research Organization (KEK), 
Tsukuba, Ibaraki 305-0801, Japan
}

\author{K.M. Kojima}
\affiliation{
Department of Physics, Graduate School of Science, University
of Tokyo, Bunkyo-ku, Tokyo 113-0033, Japan
}

\author{M. Azuma}
\affiliation{
Institute for Chemical Research, Kyoto University, Uji, Kyoto
611-0011, Japan
}
\affiliation{
PRESTO, Japan Science and Technology Corporation (JST), Kawaguchi, 
Saitama 332-0012, Japan
}
\author{M. Takano}
\affiliation{
Institute for Chemical Research, Kyoto University, Uji, Kyoto
611-0011, Japan
}

\date{\today}% It is always \today, today,
\begin{abstract}
We  report on the magnetic phase diagram of a hole-doped cuprate \cana,
 which is free from buckling of CuO$_2$ planes, determined by muon spin
 rotation and relaxation. It is characterized  by a quasi-static spin
 glass-like phase over a range of sodium concentration 
($0.05\leq x\leq 0.12$), which is held between long range
 antiferromagnetic (AF) phase ($x\leq 0.02$) and superconducting phase
 where the system is non-magnetic for $x\geq 0.15$. The obtained
 phase diagram qualitatively agrees well with that commonly found for
 hole-doped high-\tc\/ cuprates, strongly suggesting that the incomplete
 suppression of the AF order for $x>0.02$ is an essential feature of the
 hole-doped cuprates.
\end{abstract}

\pacs{74.25.Ha, 74.72.-h, 76.75.+i}% PACS, the Physics and Astronomy
\maketitle

Since the discovery of high-\tc\/ superconductivity, 
the magnetic phase diagram of layered CuO$_{2}$ materials as a function of
carrier doping has been one 
of the key issues directly related to the mechanism of superconductivity. 
It is known that while all insulating parent compounds of the planar
cuprate superconductors exhibit long range antiferromagnetic (AF) order
below a well-defined temperature (\tn), insulator-to-metal
transition and associated suppression of the AF order 
occurs when they are subjected to carrier doping. 
A copper oxychloride \cana\/ (Na-CCOC) \cite{Hiroi:94,Hiroi:96} is a
model system for studying lightly doped  states of the single CuO$_2$
plane.  This superconductor with the maximum \tc\/ = 28~K at $x\sim$ 0.2
can be derived from the prototypical high-\tc\/ superconducting
material, \lasr\/ (LSCO), by replacing lanthanum (strontium) atoms by
calcium (sodium) and the oxygen atoms at the apices of CuO$_6$ octahedra
by chlorine. Recently, it draws much attention due to their unique
character that the crystal structure is free from any distortion down to
the lowest temperature \cite{Vaknin:97}. 
Moreover, excellent cleavability of the single crystals
\cite{Kohsaka:02} enabled investigations of the electronic state in
lightly doped CuO$_{2}$ planes by surface sensitive measurements such as
angle resolved photoemission spectroscopy (ARPES) and scanning tunneling
microscopy and spectroscopy (STM/STS).

According to the results of ARPES measurements on
a superconducting specimen ($x=0.1$), the band dispersion is a consequence of the
valence band of the parent insulator shifting to the chemical
potential in accordance with the hole doping \cite{Kohsaka:03}. 
The resulting fingerprints of the parent insulator manifest
themselves in the form of a shadow band and a large pseudogap. These
results are remarkably different from the prevailing picture proposed
for the prototypical high-\tc\/ superconducting material LSCO, where
the chemical potential remains fixed while new states are created around
it by doping. It is speculated that these differences between Na-CCOC and
LSCO may be due to the difference of apical site and associated
distortion of CuO$_{2}$ planes. More interestingly, 
the real space imaging of underdoped Na-CCOC ($0.06\leq x\leq0.12$) by
the STM/STS has revealed that there are nanoscale inhomogeneity in the
electronic structure with a characteristic scale of 4--5 times the lattice
constant \cite{Kohsaka:04} and also checkerboard-like electronic
crystal, suggesting the coexistence of superconductivity and charge
ordering \cite{Hanaguri:04}; it must be noted that such measurements
for underdoped cuprates have become possible for the first time in Na-CCOC.
Meanwhile, little is known on the magnetic property of Na-CCOC
except for the parent compound ($x=0$) \cite{Vaknin:97}, since it is
hard to prepare large samples for neutron measurements and to detect the
susceptibility by SQUID. 
Thus, it would be of great interest, particularly over the underdoped region,
to clarify the ground state property of Na-CCOC in terms of magnetism.

In this Letter, we report on the magnetic phase diagram of 
Na-CCOC ($0\le x \le 0.20$) determined by muon spin rotation and
relaxation (\msr) measurements.
We have observed clear muon spin precession signals in the lightly doped
samples ($x\leq$0.02), indicating the appearance of a long range AF order. 
While no precession signal was identified in those with $0.05\leq x\leq$
0.12, evidence was found for a quasi-static spin glass(SG)-like state at
lower temperatures. 
It suggests a general trend that the suppression of long range AF order is 
strong but incomplete over the relevant range of sodium concentration. 
In particular, the present result is the first example for a good
correspondence between such an inhomogeneous magnetic state 
and the nanoscale electronic inhomogeneity revealed by STM/STS
\cite{Kohsaka:04}. 
Moreover, the obtained magnetic phase diagram 
resembles those of the hole-doped high-\tc\/ cuprates
\cite{Niedermayer:98}, supporting the universality of the present phase
diagram which is important for the basic understanding of the hole-doped
cuprates. 

The Na-CCOC compounds used in this study were synthesized under high
pressure \cite{Hiroi:94,Hiroi:96}. The samples were characterized by
means of magnetization and powder X-ray diffraction. The lattice 
parameters were determined by using a Rietveld analysis for all samples in order to 
evaluate the sodium concentration dependence of the lattice parameters. 
%In zero field (ZF)-\msr\/ experiment, a beam of nearly 100\% polarized
%muons is focused on a target sample. After stopping almost
%instantaneously at interstitial sites, each muon precesses in the
%internal field $B$ at a frequency $2\pi f = \gamma_\mu B$ 
%($\gamma_\mu = 135.54$ MHz/T). When the muon decays, the decay positron
%is emitted preferentially along the muon spin direction. As a result, the
%accumulated positron time histograms allow one to monitor the muon spin
%evolution. 
Zero field (ZF)-\msr\/ measurements were conducted at TRIUMF
and at the Muon Science Laboratory, High Energy Accelerator Research Organization
(KEK-MSL). We prepared eleven sets of Na-CCOC polycrystalline specimens, 
$x=$ 0, 0.0025, 0.005, 0.01, 0.02, 0.05, 0.07, 0.10, 0.12, 0.15, and 0.20, 
having a dimension of about 100 $\sim$ 250 mm$^2$ with $\sim$ 1 mm thickness.
These samples were mounted on a thick sample holder (KEK-MSL) or on a
thin sheet of mylar film (TRIUMF, where one can obtain background free
spectra) and loaded to the $^4$He gas flow cryostat. 
ZF-\msr\/ measurements were mainly performed at temperatures between
2~K and room temperature, and additional measurements
were performed at ambient temperature under a transverse field 
($\simeq 2$~mT) to calibrate the instrumental asymmetry.
The dynamics of local magnetic fields at the muon sites was 
investigated by longitudinal field (LF)-\msr\/ measurements 
\cite{Schenck:86}.

\begin{figure}[tb]
\includegraphics[width=0.8\linewidth]{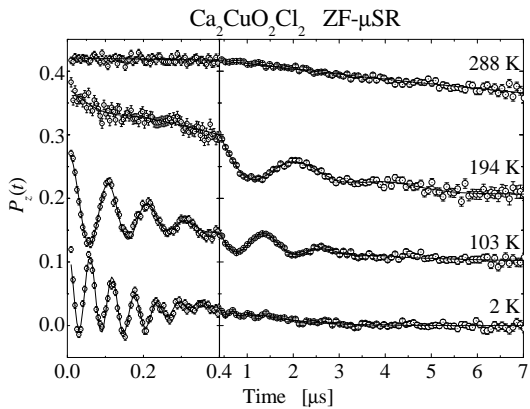}% Here is how to import EPS art
\caption{\label{precession} ZF-\msr\/ time spectra, $P_z(t)$, in CCOC
at various temperatures. The left and
 right column corresponds to $P_z(t)$ for 0--0.4 $\mu$s and 0.4--7 $\mu$s,
 respectively. Note that $P_z(t)$ values of 288 K and 194 K are added
 0.2, and that of 103 K is added 0.1.}
\end{figure}
\begin{figure}[tb]
\includegraphics[width=0.8\linewidth]{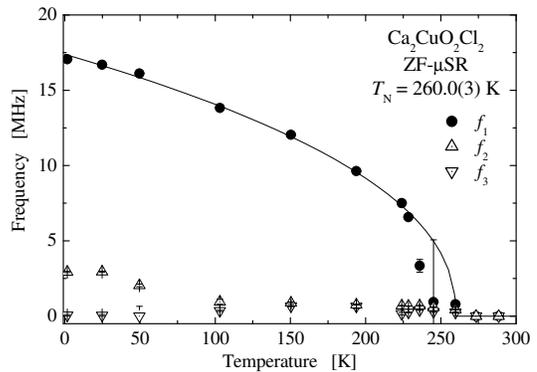}% Here is how to import EPS art
\caption{\label{frq} Temperature dependence of the muon spin precession
frequencies in CCOC. 
Solid curve is a fitting result for the data $f_1$
by a form $A(T_{\rm N}-T)^\gamma$,
 where \tn\/ = 260.0(3) K is obtained.}
\end{figure}
Figure \ref{precession} shows the time spectra of the muon spin polarization
$P_z(t)$ observed in non-doped CCOC under zero field at typical temperatures, 
where muon spin precession signals are
clearly observed below \tn\/ $\simeq$ 260~K. 
These signals unambiguously demonstrate that the 
system falls into a long range AF state below \tn. 
Consequently, the data were analyzed
by the following function; 
$
%\begin{equation}
P_z(t)=\sum^n_{i=1}A_i\exp[-(\sigma_it)^2]\cos(2\pi f_it+\phi)+A_{\rm n}\exp[-(\sigma_{\rm n}t)^\beta],
\label{fnc1}
%\end{equation}
$
where $A_i$ and $A_{\rm n}$ are the asymmetry of oscillating and
non-oscillating components, $f_i$ is the muon spin precession frequency,
$\sigma_i$ and $\sigma_{\rm n}$ are the relaxation rate, $\beta$ is
the power of the exponent. The relevant spectra taken at TRIUMF
were free from background.

The temperature dependence of $f_i$ is shown
in Fig.~\ref{frq}.  It exhibits a sharp reduction towards 260~K, where the
oscillating component disappears around 275~K. This is consistent with
the previous neutron result in which Bragg peak appears below 
\tn\/ = 247(5)~K \cite{Vaknin:97}. 
As shown in Fig.~\ref{precession} and Fig.~\ref{frq}, two different
frequencies are observed below 260~K, and one of which exhibits further
splitting into two components below around 100~K. 
It is interesting to note that one of 
those components ($f_2$) takes a value common to that in LSCO ($\sim$ 5~MHz)
\cite{Uemura:87,Uemura:88,Uemura:89}, and $f_1$ seems to be common
to that in \sr\/ (SCOC) \cite{Le:90}. The data of $f_1$ are well
reproduced by a function of the form $A(T_{\rm N}-T)^\gamma$, with 
\tn\/ = 260.0(3)~K and $\gamma=$ 0.44(1). 

While clear muon spin precession signals were observed in those with
small $x$, \tn\/ showed a steep decrease with increasing sodium concentration. 
A fit procedure similar to that for CCOC has been applied to the data in
the Na-doped samples ($0 < x\leq$ 0.02). 
The splitting of frequency from two to three components below
around 100~K were commonly observed in the samples with $x$ = 0.0025
(\tn\/ = 240.6(8)~K) and 0.005 (\tn\/ = 185.3(1.3)~K). On the other
hand, there remained only two components below \tn\/ in $x$ = 0.01
(\tn\/ = 34.5(3)K) and 0.02 (\tn\/ = 6.6~K).

\begin{figure}[tb]
\includegraphics[width=0.8\linewidth]{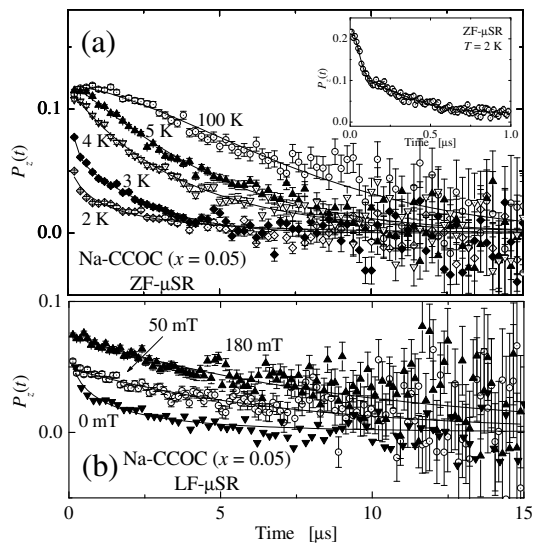}% Here is how to import EPS art
\caption{\label{05time}(a) ZF-\msr\/ time spectra 
in Na-CCOC with $x=0.05$ at various temperatures. The inset shows a
 ZF-\msr\/ time spectrum at 2~K where no spontaneous precession signal
 is observed. (b) LF-\msr\/ time spectra at 2~K under various applied
 fields. $P_{z}(t)$ was obtained by subtracting the
 temperature-independent contribution from the silver sample holder.}
\end{figure}
Figure \ref{05time} shows the ZF-\msr\/ time spectra in a specimen
with $x$ = 0.05 at various temperatures. As shown in the inset, no muon
spin precession signal was observed at the lowest temperature ($T$ = 2~K). 
It clearly indicates that a strongly disordered magnetic ground state dominates
over the AF long range order at this sodium concentration. 
The depolarization is well described by
$
%\begin{equation}
P_z(t)=A_1\exp(-\lambda_1 t)\exp[-(\Delta t)^2]+A_2\exp(-\lambda_2 t)+A_{\rm B},
\label{05fnc}
%\end{equation}
$
where $A_i$ is the decay asymmetry for signals
from the sample, $A_{\rm B}$ is that from the sample holder, 
$\lambda_i$ is the relaxation rate, 
and $\Delta$ is the {\it nuclear} dipolar width predominantly determined
by $^{63,65}$Cu and $^{35,37}$Cl nuclear moments (see below).
The values of $\Delta$ were fixed to those determined at ambient
temperature where $\lambda_i\simeq0$. The second term is needed only
below 3~K. Solid curves in Fig.~\ref{05time}(a) are the fit results. 
%by Eq.~(\ref{05fnc}). 
It was revealed that while $\lambda_1$ is almost independent
of temperature above 6~K, it starts to increase rapidly with decreasing
temperature below 6~K. 
Such fast relaxation without any detectable oscillation is a common 
signature of disordered magnetism which is either static or dynamically
fluctuating. The two possibilities can be discriminated by the field dependence
of $\lambda_i$ under a LF. 
Figure~\ref{05time}(b) shows the LF-\msr\/ time spectra at
2~K under various magnetic fields ($x=0.05$). It is
clear from the analysis that considering the effect of LF, the
relaxation rate remains almost unchanged with increasing field from 50
to 180 mT, strongly suggesting that the local field probed by muons is
fluctuating due to the slowing down of electronic Cu moments. 
The fluctuation rate becomes slow enough below 6~K where $\lambda_1$
starts to increase, where 
$\lambda_1\simeq (\gamma_\mu A_\mu)^2/\nu\sim10^5$ s$^{-1}$
with $\gamma_\mu A_\mu\simeq 2\pi f_1(T\rightarrow0)$. Here, 
$\gamma_\mu$ is the muon gyromagnetic ratio ($=2\pi\times$135.54 MHz/T). 
This leads to an estimation for the fluctuation rate, $\nu\sim10^9$ s$^{-1}$
at 2~K. We made additional measurements using a $^3$He--$^4$He dilution
refrigerator in the specimen with $x=0.10$
and found that a quasi-static SG-like state is realized
below $\sim0.6$~K. This indicates that the increase of $\lambda_i$ is a
precursor of a SG-like transition; the temperature region over which
the fluctuation is observed is much wider than that associated with the
AF transition for $x\leq 0.02$. Thus, we define the
onset of SG-like behavior at \td\/ $\sim$ 6~K in $x=0.05$
specimen. A similar behavior in both ZF-\msr\/ and LF-\msr\/ was observed 
for $x=0.07$, 0.10 and 0.12, where $\lambda_1$ increases below \td\/
$\sim$ 4.3~K, 3.5~K and 2.5~K, respectively. 

\begin{figure}[tb]
\includegraphics[width=0.8\linewidth]{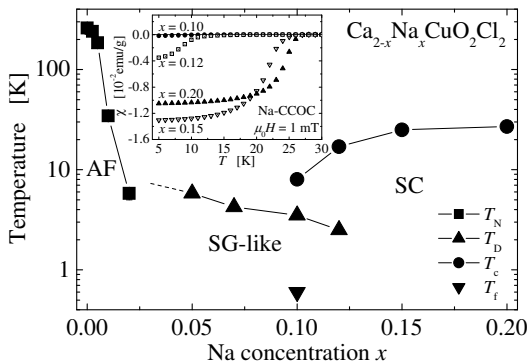}% Here is how to import EPS art
\caption{\label{diagram} Magnetic phase diagram of Na-CCOC. 
The transition temperature for the SG-like state
(\td) is defined as an onset temperature where the spin
fluctuation rate becomes lower than $10^9$ s$^{-1}$. 
The spin freezing temperature ($T_{\rm f}$) is only defined in $x=0.10$ specimen.
The inset shows the temperature dependence of magnetic susceptibility measured for
($x=0.10$, 0.12, 0.15 and 0.20) in an applied field $\mu_0H=1$~mT
 after field-cooling process.}
\end{figure}

The superconducting character of the present samples is
examined by the temperature dependence of the magnetic
susceptibility, as shown in the inset of Fig.~\ref{diagram} for 
$x=0.10$, 0.12, 0.15 and 0.20.
A clear sign of the Meissner effect 
is observed in the samples with $x=0.12,0.15$ and 0.20, 
while no bulk superconductivity is
detected for $x=0.10$.  From these data, the superconducting 
volume fraction was estimated to be 1\%, 
16\%, 60\% and 48\% in the respective samples with $x=$ 0.10,
0.12, 0.15 and 0.20. 
Since the volume fraction is very small in those
with $x=0.10$ and 0.12, it is reasonable to presume 
a phase separation in those samples with the superconducting phase as a minority
phase.  

ZF-\msr\/ time spectra in the specimen with $x=0.15$ and 0.20
exhibit the least temperature dependence down to 2~K. 
The relaxation is predominantly due to 
static and randomly oriented nuclear magnetic
moments which give rise to depolarization described by 
the Kubo-Toyabe function, $G_{\rm KT}(t)$. It is
characterized by a Gaussian decay, $\sim\exp[-(\Delta t)^2]$, at early
times followed by a recovery of the asymmetry to 1/3, 
%\begin{equation}
%G_{\rm KT}(t)=\frac1{3}+\frac2{3}\left(1-\Delta^2t^2\right)\exp\left(-\frac1{2}\Delta^2t^2\right),
%\label{KT}
%\end{equation}
where $\Delta$ is the rms width of the field distribution arising from
the nuclear moments. 
Besides the signature of these static nuclear
magnetic moments, we obtained no evidence for any kind of additional
magnetic moments in the specimen with $x=0.15$ and 0.20. 
The only exception is that the spectrum obtained at 2~K
seems to exhibit a tiny enhancement of relaxation compared with that at 171~K. 
Considering this point, the data were analyzed by the following function, 
$
%\begin{equation}
P_z(t)=AG_{\rm KT}(t)\exp(-\Lambda t)+A_{\rm B}
\label{15fnc}
%\end{equation}
$
, which yields $\Delta$ = 0.154(20) $\mu$s$^{-1}$ ($x=0.15$) and 
0.162(8) $\mu$s$^{-1}$ ($x=0.20$). 
Although $\Lambda$ ($\sim10^{-2}$ $\mu$s$^{-1}$) slightly 
depends on temperature, it may be attributed to a slow dynamics of
muon (e.g., diffusion). Another possibility is that $P_z(t)$ may be a sum 
of $G_{\rm KT}(t)$ with different values for $\Delta$ due to 
multiple muon stopping sites with varying distance to the Cu and Cl nuclear
moments (see below). As a summary of our results, the magnetic phase
diagram of Na-CCOC as a function of $x$ is shown in Fig.~\ref{diagram}. 

Since the electronic Cu moment is the only source of strong local
magnetic fields ($\gg 1$ mT) in the AF phase,
the observation of multiple components in the precession signals 
for $x\leq 0.02$ suggests that there are multiple
muon sites in Na-CCOC. In the case of ZF-\msr, muons experience a local
field $\mbox{\boldmath $H$}_{loc}=\mbox{\boldmath
$H$}_{dip}+\mbox{\boldmath $H$}_{fc}$, where $\mbox{\boldmath $H$}_{dip}=\sum_i(3r^\alpha_i r^\beta_i/r^2_i-\delta_{\alpha\beta})/r^3_i$ ($\alpha,\beta=x,y,z$)
is the magnetic dipolar field from the $i$-th Cu$^{2+}$ ion at a distance
$r_i$ from the muon, and
$\mbox{\boldmath $H$}_{fc}$ is the Fermi-contact hyperfine field 
due to the finite electron spin density at the muon site. Since the latter
is likely to be negligible as in the case of \la\/ (LCO), 
we can calculate the local field 
at a muon site as the sum of magnetic dipolar field from the Cu moments
with their moment size $\simeq 0.25(10)\mu_{\rm B}$ 
as determined by neutron diffraction \cite{Vaknin:97}. 
 Another source of information is the nuclear dipolar
width $\Delta$ determined in the paramagnetic state.
In the case of polycrystalline samples, $\Delta$ 
is determined by the second moment, 
%\begin{equation}
$
\Delta^2=\frac5{3}\gamma_{\mu}^2\overline{\sigma_{\rm VV}^2}=\frac4{9}\gamma_{\mu}^2\sum_i\overline{\mu_i^2}\frac1{r_i^6},
\label{Delta}
$
%\end{equation}
where $\sigma_{\rm VV}$ is a value defined by Van Vleck \cite{Hayano:79}
and $\mu_i$ is the {\it nuclear} magnetic moment situated at distance $r_i$ from the
muon site.  Comparison of the experimental values of $H_{\rm dip}$ and $\Delta$
with those calculated by the above formulae yields
$R_1=1.97$\AA\/ (from $f_1$) and $R_2=3.52$\AA\/ (from $f_2$), respectively. 
Considering the distance of $r_{\rm Cu-O}=1.934$\AA\/ in CuO$_2$ plane and 
$r_{\rm Cu-Cl}=2.737$\AA\/ in CCOC, we conclude that $f_1$ and $f_2$ 
respectively correspond
to the muons stopped near the planer oxygens and apical chlorines. 
The latter site is close to that in LCO, which is perfectly in line with the
fact that $f_2$ is close to that observed in LCO. 
The above conclusion is also consistent with a theoretical estimation made for
LCO \cite{Sulaiman:94}, where muons are predicted to
be attracted by the negative charge at the apical oxygen atoms.
The appearance of additional site for CCOC may be explained by the 
reduced ionic charge of Cl$^-$, which makes
the site near the apical Cl less attractive to muons compared with that near
O$^{2-}$ in terms of the electrostatic energy. Then it seems reasonable
that a considerable fraction of implanted muons dwells on the CuO$_2$ plane 
containing O$^{2-}$ ions. 

The revelation of SG-like phase in the underdoped Na-CCOC is one
of the most important results in the present study.
As mentioned earlier, the very recent STM/STS measurements on the 
corresponding specimen have revealed a nanoscale inhomogeneity in the
electronic density of state (DOS) \cite{Kohsaka:04}. Considering that
the SG-like state is observed as a bulk majority over the region
$0.05\leq x\leq 0.12$, it is natural to presume that the quasi-static
SG-like state (magnetic inhomogeneity) has a close link with the
electronic inhomogeneity observed in the underdoped Na-CCOC. 
However, the possibility to interpret the inhomogeneity as those
consisting of the AF domains is ruled out; it would lead to the
spontaneous muon precession with an amplitude proportional to the volume
fraction of AF domains \cite{Savici:02}. Accordingly, the low DOS area
probed by STS/STM should not be interpreted simply as the AF
domains. The picture of diluted local moments would be also inconsistent
with the STM/STS observation, since such moments would not induce the
nanoscale inhomogeneity. Thus, one of the few possibilities to explain
both \msr\/ and STM/STS results would be to consider a spin density wave
(SDW) associated with the DOS distribution. 
We point out that the observed time spectrum shown in the inset of in 
Fig.~\ref{05time}(a) exhibits a close resemblance with that in the SDW
state \cite{Ohishi:00}. The sensitivity to induce
such an inhomogeneity upon the hole-doping would be a key to understand
the electronic state of CuO$_2$ planes. We also note that no strong
anomaly was observed for $x=0.12$ ($\simeq1/8$), suggesting either the
buckling of CuO$_2$ planes or the ionic radius of alkaline metals
(Ca$<$Sr$<$Ba) may be related to the 1/8 anomaly. 

The remaining issue is the splitting of frequency below
100~K. In the case of SCOC, it is also reported that the muon
precession frequency splits into two components below 60~K \cite{Le:90},
which is attributed to an intrinsic effect of possible spin reorientation. 
Accordingly, one possible scenario is that a similar spin reorientation
may occur also in CCOC. However, it turns out
that both neutron and magnetization measurements report no sign 
of such spin-flop transition in CCOC \cite{Vaknin:97}.  
The fact that $f_1$ remains unique below 100~K also disfavors this scenario.
Another possibility is that muons near the apical Cl ions
may undergo a local site change below 100~K, although the origin of 
the instability is unclear at this stage.

%In summary, our \msr\/ measurements on Na-CCOC revealed a magnetic
%phase diagram as a function of carrier doping $x$, which shares many 
%features in common with other hole-doped cuprates. 
%The AF order observed in the sample with 
%$0\leq x \leq 0.02$ is drastically
%suppressed in those with $x>0.02$. However, the suppression is
%incomplete so that a SG-like phase appears over a region
%$0.05\leq x \leq 0.12$. The present result strongly suggests that
%the inhomogeneous magnetic state commonly found in the underdoped
%cuprates has an intrinsic origin related to the nanoscale electronic inhomogeneity 
%observed by STM/STS measurements.

We would like to thank the staff of TRIUMF and KEK for their technical
support during the experiments. This work was partially supported by a
Grant-in-Aid for Creative Scientific Research and a Grant-in-Aid for
Scientific Research on Priority Areas by Ministry of Education,
Culture, Sports, Science and Technology, Japan.


\end{document}